\documentclass[12pt]{article}
\usepackage{graphics}
\usepackage{cite}
\usepackage{epsfig}
\usepackage{latexsym}
\newcommand  {\etal}     {{\it et al.}}

\newcommand  {\BJ}       {{\it Biophys.\ J.\ }}

\newcommand  {\COSB}     {{\it Curr.\ Opin.\ Struct.\ Biol.\ }}

\newcommand  {\EL}       {{\it Europhys.\ Lett.\ }}

\newcommand  {\JACS}     {{\it J.\ Am.\ Chem.\ Soc.\ }}

\newcommand  {\JCP}      {{\it J.\ Chem.\ Phys.\ }}
\newcommand  {\JMB}      {{\it J.\ Mol.\ Biol.\ }}

\newcommand  {\MP}       {{\it Mol.\ Phys.\ }}

\newcommand  {\Nat}      {{\it Nature\ }}
\newcommand  {\NSB}      {{\it Nat.\ Struct.\ Biol.\ }}

\newcommand  {\Pro}      {{\it Proteins\ }}

\newcommand  {\ProSci}   {{\it Protein\ Sci.\ }}

\newcommand  {\PNAS}     {{\it Proc.\ Natl.\ Acad.\ Sci.\ USA\ }}

\newcommand  {\Sci}      {{\it Science\ }}

\newcommand{\beq}{\begin{equation}}
\newcommand{\eeq}{\end{equation}}
\newcommand{\beqa}{\begin{eqnarray}}
\newcommand{\eeqa}{\end{eqnarray}}
\newcommand{\bea}{\begin{eqnarray}}
\newcommand{\eea}{\end{eqnarray}}

\newcommand   {\Fs}      {F${}_{\mbox{{\small s}}}$}
\newcommand   {\lop}     {L\'opez de la Paz}
\newcommand   {\Nhb}     {N_{\mbox{{\scriptsize hb}}}^{\mbox{{\scriptsize nat}}}}
\newcommand   {\Ca}      {C${}_{\alpha}$}
\newcommand   {\Cb}      {C${}_{\beta}$}

\newcommand   {\Cp}      {C${}^{\prime}$}
\newcommand   {\dE}      {\Delta E}

\newcommand   {\Eev}     {E_{\mbox{{\scriptsize ev}}}}
\newcommand   {\Eloc}    {E_{\mbox{{\scriptsize loc}}}}
\newcommand   {\Ehb}     {E_{\mbox{{\scriptsize hb}}}}
\newcommand   {\Ehp}     {E_{\mbox{{\scriptsize hp}}}}
\newcommand   {\Ehpu}    {E_{\mbox{{\scriptsize hp}}}^{\mbox{{\scriptsize u}}}}
\newcommand   {\Ehpn}    {E_{\mbox{{\scriptsize hp}}}^{\mbox{{\scriptsize n}}}}
\newcommand   {\Tm}      {T_{\mbox{{\scriptsize m}}}}

\newcommand   {\Pn}      {P^{\mbox{{\scriptsize n}}}}
\newcommand   {\Pu}      {P^{\mbox{{\scriptsize u}}}}

\newcommand   {\rc}      {r^{\mbox{{\scriptsize c}}}}

\newcommand   {\kev}     {\kappa_{\mbox{{\scriptsize ev}}}}
\newcommand   {\kloc}    {\kappa_{\mbox{{\scriptsize loc}}}}

\newcommand   {\ehba}    {\epsilon^{(1)}_{\mbox{{\scriptsize hb}}}}
\newcommand   {\ehbb}    {\epsilon^{(2)}_{\mbox{{\scriptsize hb}}}}

\newcommand   {\Mij}     {M_{IJ}}

\begin{document}

\begin{flushright}
LU TP 03-40\\
November 13, 2003
\end{flushright}
 
\vspace{0.4in}
 
\begin{center}
 
{\LARGE \bf Folding Thermodynamics of Three
$\beta$-Sheet Peptides: A Model Study}
 
\vspace{.6in}
 
\large
Anders Irb\"ack and Fredrik Sjunnesson\footnote{E-mail: anders,\,fredriks@thep.lu.se}\\
\vspace{0.10in}
Complex Systems Division, Department of Theoretical Physics\\
Lund University,  S\"olvegatan 14A,  SE-223 62 Lund, Sweden \\
{\tt http://www.thep.lu.se/complex/}\\
 
\vspace{0.3in}
 
Submitted to \Pro
 
\end{center}
\vspace{0.3in}
\normalsize
Abstract:\\
We study the folding thermodynamics of a $\beta$-hairpin and two
three-stranded $\beta$-sheet peptides using a simplified
sequence-based all-atom model, in which folding is driven
mainly by backbone hydrogen bonding and effective hydrophobic
attraction. The native populations obtained for these three
sequences are in good agreement with experimental data. We also
show that the apparent native population depends on which
observable is studied; the hydrophobicity energy and the number of native
hydrogen bonds give different results. The magnitude of this
dependence matches well with the results obtained in two
different experiments on the $\beta$-hairpin.

\newpage

\section{Introduction}

Peptide folding is currently attracting considerable attention.  
Recent advances in this area include the {\it de novo} design of 
two monomeric three-stranded antiparallel $\beta$-sheet peptides,  
Betanova\cite{Kortemme:98,Lopez:01} and Beta3s.\cite{deAlba:99} 
Peptides that have the ability to fold on their own and are well 
characterized experimentally are valuable not least as a testbed for 
theoretical models for protein folding. $\beta$-sheet 
peptides are particularly interesting in this respect, as
$\beta$-sheet formation is more challenging to model than
$\alpha$-helix formation. Therefore, it is no surprise that both 
Betanova\cite{Bursulaya:99,Colombo:02} and Beta3s\cite{Cavalli:03} 
have become the subject of computational studies. Simulations
of peptide sequences that are somewhat similar to these and occur in 
natural proteins, so-called WW domains, have been reported, 
too.\cite{Karanicolas:03} For a recent review of computational
studies of peptide folding, see Granakaran~\etal\cite{Granakaran:03}   

Here we present a study of the C-terminal $\beta$-hairpin 
from the protein G B1 domain and a triple mutant 
of Betanova called LLM.\cite{Lopez:01} The original 
Betanova, which is less stable 
than the peptide LLM,\cite{Lopez:01} is considered too.
These different sequences are studied 
using an all-atom model with a simplified 
interaction potential. An earlier version of this model
was tested\cite{Irback:03} on the same $\beta$-hairpin and an 
$\alpha$-helix, the designed so-called 
\Fs.\cite{Lockhart:92,Lockhart:93} 
The model was able to fold these two sequences and the folded population
showed, in both cases, a temperature dependence comparable with 
experimental data. It should be pointed out that  
different sequences are studied using exactly the same parameters; 
the interaction potential is, like 
that of Kussell~\etal\cite{Kussell:02} but unlike many other 
simplified potentials for protein folding, entirely sequence-based. This
is of importance even if only one sequence is studied, because it ensures
that the formation and breaking of non-native bonds is not a neglected part 
of the dynamics.

\section{Materials and Methods}

The model we study is a revised version of an earlier 
model.\cite{Irback:03} It contains all atoms of the polypeptide
chain, including hydrogens, but no explicit water molecules. All 
bond lengths, bond angles and peptide torsion angles 
($180^\circ$) are held fixed, so each amino acid has 
the Ramachandran torsion angles $\phi$, $\psi$ and a number of 
side-chain torsion angles as its degrees of freedom 
(for Pro, $\phi$ is held fixed at $-65^\circ$). All bond lengths and
bond angles are the same as in the original model.\cite{Irback:03}

The potential function
\begin{equation}
  E=\Eev+\Eloc+\Ehp+\Ehb
  \label{energy}
\end{equation}
is composed of four terms. The remaining part of this section describes 
these different terms, with emphasis on what is new compared with the 
earlier model. Energy parameters are quoted in dimensionless units.
To set the energy scale of the model, we use the midpoint temperature 
for the $\beta$-hairpin as determined by Mu\~noz~\etal,\cite{Munoz:97}
$\Tm=297$\,K, which corresponds to $kT\approx0.440$ in the model.

The first term in Eq.~\ref{energy}, $\Eev$, represents excluded-volume effects 
and has the form 
\beq
\Eev=\kev \sum_{i<j}
\biggl[\frac{\lambda_{ij}(\sigma_i+\sigma_j)}{r_{ij}}\biggr]^{12}\,,
\label{ev}\eeq
where $\kev=0.10$ and $\sigma_i=1.77$, 1.75, 1.55, 1.42 and 1.00\,\AA\ for
S, C, N, O and H atoms, respectively. 
The role of the parameter $\lambda_{ij}$ is to reduce the repulsion 
between non-local pairs; $\lambda_{ij}=1$ for all pairs
connected by three covalent bonds and $\lambda_{ij}=0.75$ otherwise. 
The reason for using $\lambda_{ij}<1$ for non-local pairs is partly 
computational efficiency, and partly the restricted flexibility of chains 
with only torsional degrees of freedom. To speed up the calculations,
the sum in Eq.~\ref{ev} is evaluated using a cutoff of 
$\rc_{ij}=4.3 \lambda_{ij}$\,\AA.

The second interaction term, $\Eloc$, is new compared with the earlier model.
By introducing this term and modifying $\sigma_i$ for C and N, we slightly  
adjusted the shape of the Ramachandran $\phi$, $\psi$ distribution. $\Eloc$ 
is a local electrostatic energy given by  
\beq
\Eloc=\kloc \sum_I 
\rho_I\left(\sum \frac{q_iq_j}{r_{ij}^{(I)}/{\rm \AA}}\right)\,,
\label{loc}\eeq
where the outer sum runs over all non-Pro amino acids along the chain, 
and the inner sum represents the interaction between the partial charges
of the backbone NH and \Cp O groups within one amino acid (the sum has 
four terms: N\Cp, NO, H\Cp\ and HO).
The partial charges are $q_i=\pm 0.20$ for H and N 
and $q_i=\pm 0.42$ for \Cp\ and O.\cite{Branden:91} We put $\kloc=125$, 
which corresponds to a dielectric constant of $\epsilon_r\approx 2.0$
if $\rho_I=1$. The factor $\rho_I$ reduces the interaction strength
for the two end amino acids and Gly, which can be viewed as a   
simple form of context dependence; $\rho_I=0.25$ for end amino acids, 
$\rho_I=0.5$ for Gly, and $\rho_I=1$ otherwise. A similar factor is 
used for $\Ehb$ (see below).  

The third term in Eq.~\ref{energy}, $\Ehp$, is an effective attraction 
between hydrophobic side chains that are not nearest or next-nearest
neighbors along the chain. It has the pairwise additive form  
\beq
\Ehp=-\sum\Mij C_{IJ}\,,
\label{hp}\eeq
where $C_{IJ}$ is a measure of the degree of contact between 
side chains $I$ and $J$, and $\Mij$ sets the energy that a pair
in contact gets. The contact measure $C_{IJ}$ is a number between 0 and 1,
defined as before.\cite{Irback:03} The interaction matrix $\Mij$ is given 
in Table~I and differs from that used in our earlier study, 
which was based on the Miyazawa-Jernigan contact 
energies.\cite{Miyazawa:96} With an all-atom representation, this cannot 
be expected to be  a good choice for more general 
sequences, since the Miyazawa-Jernigan contact energies 
were derived using a different, reduced chain 
representation.\cite{Miyazawa:96} 
The new matrix $\Mij$ 
has a simplified structure 
in that the hydrophobic amino acids are grouped into 
three classes (see Table~I). 
The $\Mij$ values are taken to be 
large for the aromatic class (Phe, Trp, Tyr), which in part is an 
attempt to compensate for the fact that it is relatively difficult 
for these large side chains with few degrees of freedom to make 
proper contacts. 

\begin{table}[t]
\begin{center}
\begin{tabular}{rlccc}
                 &  & I   & II & III \\
\hline
I& Ala                & 0.0   & 0.1  & 0.1   \\
II& Ile, Leu, Met, Val &       & 0.9  & 2.8   \\
III& Phe, Trp, Tyr      &       &      & 3.2         
\end{tabular}
\caption{
The interaction matrix $\Mij$ (see Eq.~\protect\ref{hp}). 
All amino acid pairs not occurring in the table have $\Mij=0$.}
\label{tab:1}
\end{center}
\end{table}

The last term of the potential, the hydrogen-bond energy $\Ehb$, 
is given by  
\begin{equation}
  \Ehb= \ehba \sum_{{\rm bb-bb}}
  \rho_{ij}u(r_{ij})v(\alpha_{ij},\beta_{ij}) +
  \ehbb \sum_{{\rm sc-bb}} 
  \rho_{ij}u(r_{ij})v(\alpha_{ij},\beta_{ij})\,, 
  \label{hbonds}
\end{equation}
where the two terms represent backbone-backbone interactions and
interactions between the backbone and charged side chains, respectively. 
The second term in Eq.~\ref{hbonds} does not include any side chain-side
chain interactions, as it did in our earlier study. 
Apart from that,
the only difference compared with the earlier model is the factor
$\rho_{ij}$, which like $\rho_I$ in Eq.~\ref{loc} can be seen as a
simple form of context dependence.  We put $\rho_{ij}=0.25$ if any of
the two amino acids involved is an end amino acid, $\rho_{ij}=0.5$ if
any of them is a Gly, and $\rho_{ij}=1$ otherwise. The constants
$\ehba=3.1$ and $\ehbb=2.0$ as well as the functions $u$ and $v$ are
exactly the same as before.\cite{Irback:03}

To study the thermodynamic behavior of this model, we use 
simulated tempering,\cite{Lyubartsev:92,Marinari:92}
in which the temperature is a dynamical variable. Details on our
implementation of this method can be found elsewhere.\cite{Irback:95}
For a review of simulated tempering and other generalized-ensemble 
techniques for protein folding, see Hansmann and Okamoto.\cite{Hansmann:99}
Eight different temperatures are studied, ranging from 284\,K to 371\,K.  
For the backbone degrees of freedom, we use three different elementary
moves: first, the pivot move\cite{Lal:69} in which a single
torsion angle is turned; second, a semi-local method\cite{Favrin:01} that
works with up to eight adjacent torsion angles, which are turned in a
coordinated way; and third, a symmetry-based update of
three randomly chosen backbone torsion angles.\cite{Irback:03} 
For the side-chain degrees of freedom, we use simple 
Metropolis updates of individual angles.  

For each peptide, eight independent Monte Carlo runs were performed, 
starting from random conformations. Each run 
required a few days on a standard desktop computer, and 
contained several folding/unfolding events. The similarity between
the results from the different runs strongly suggest that the 
simulations did map out all relevant free-energy minima of the model. 
All statistical errors quoted are 1$\sigma$ 
errors obtained from the variance between the runs. 
The fits of data discussed in the next section were carried 
out by using a Levenberg-Marquardt procedure.\cite{NR}

For a given protein structure, there generally exist alternative structures
with similar secondary-structure content but different overall 
topologies. This holds true even for a small $\beta$-hairpin, for which
a flip of the side chains gives rise to a topologically distinct structure.
To make models discriminate between different topologies is a delicate 
task. To assess whether or not a model is able to do that, it is necessary 
to make a suitable choice of observables. In our calculations, 
we monitor two variables that can be used for this purpose:
first, the root-mean-square deviation (rmsd) from the folded 
structure, $\Delta$, calculated over all non-H atoms 
(a backbone rmsd is much less informative); and second, the number of 
native backbone-backbone hydrogen bonds, $\Nhb$. Figure~1 
illustrates which hydrogen bonds we take to be present in the native 
states of the peptides studied. 
\begin{figure}[t]
\begin{center}
\epsfig{figure=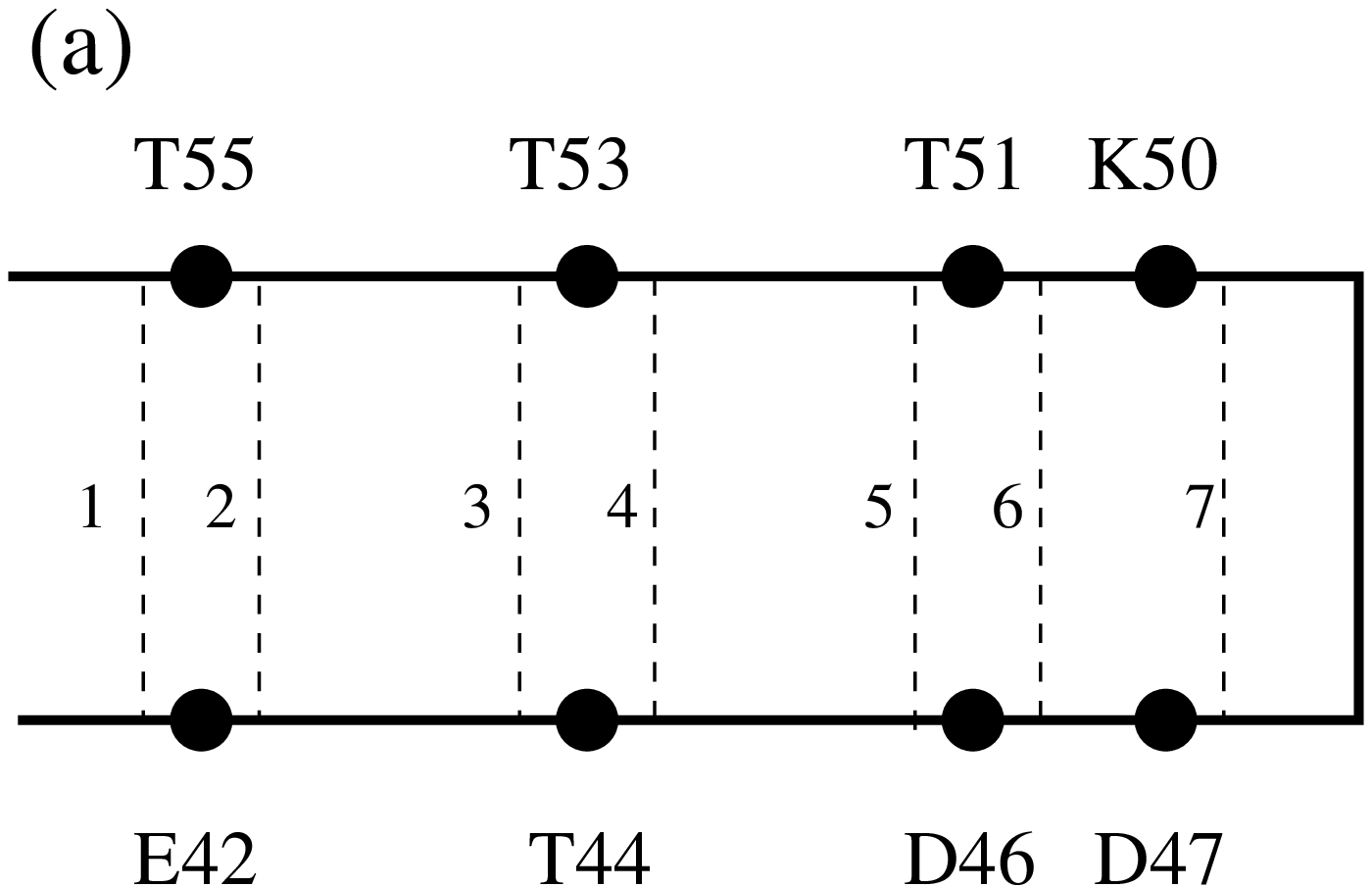,width=7cm}
\hspace{5mm}
\epsfig{figure=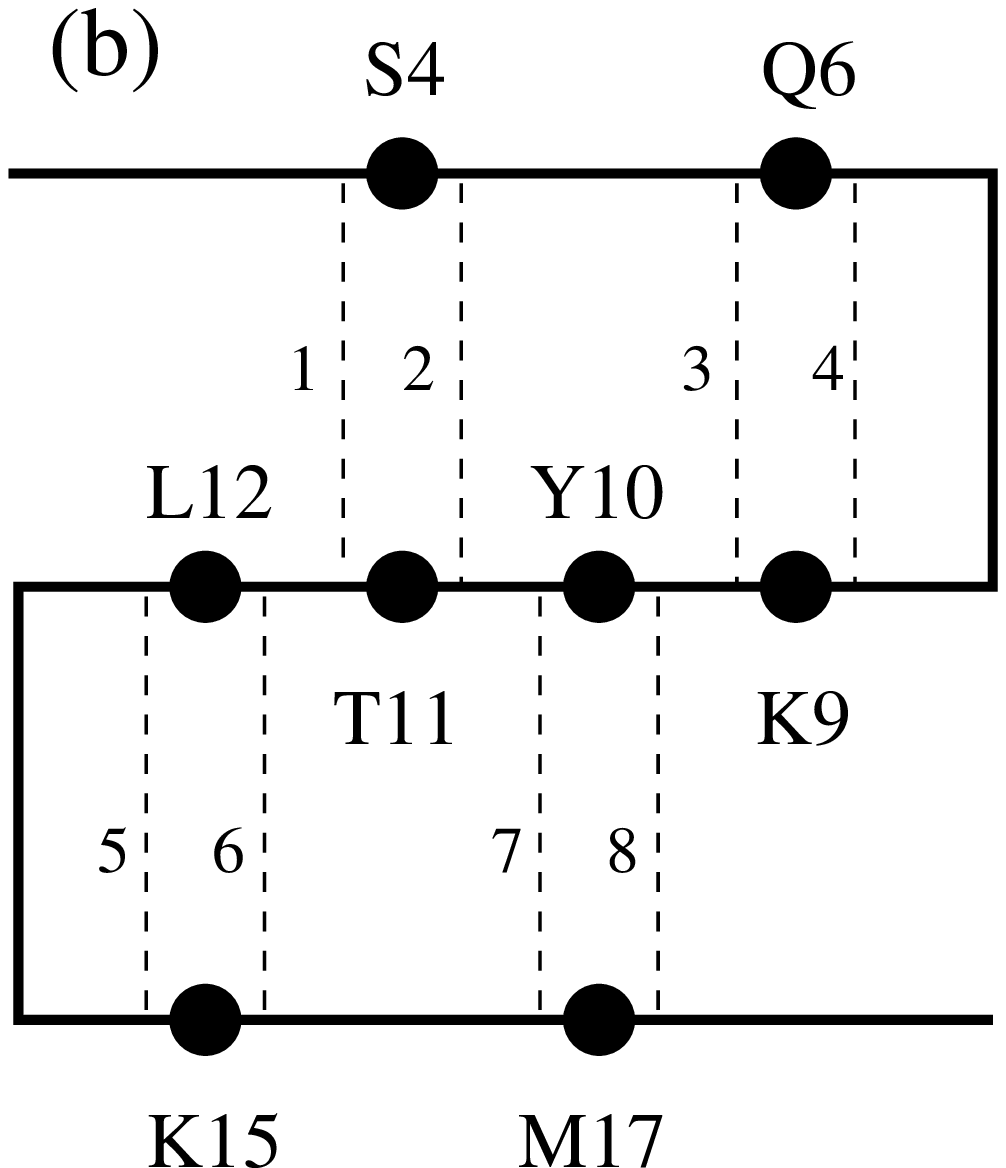,width=4.7cm}
\end{center}
\vspace{-3mm}
\caption{
Schematic illustration of the 
backbone-backbone hydrogen bonds taken as native for
(a) the C-terminal $\beta$-hairpin from the protein G B1 
domain,\cite{Blanco:94,Munoz:97} and (b) the mutant LLM of
Betanova.\cite{Lopez:01} Diagram (b) is used for the original Betanova, 
too (with L12 and M17 replaced by N12 and T17, respectively).}
\label{fig:1}\end{figure}
In our calculations, a hydrogen bond 
is considered formed if the energy is less than $-\ehba/3$ 
(see Eq.~\ref{hbonds}).

Using the original model, we studied the $\alpha$-helical \Fs\ peptide and 
a $\beta$-hairpin.\cite{Irback:03} Here, we study the
same $\beta$-hairpin and two three-stranded $\beta$-sheet 
peptides, LLM and Betanova. Before turning to these results, it should 
be pointed out that the \Fs\ sequence still makes an $\alpha$-helix 
in the revised model, as can be seen from the free energy $F(\Delta,E)$ 
in Fig.~2. $F(\Delta,E)$ has a pronounced, dominating 
minimum at $\Delta\approx2$--3.5\,\AA, which corresponds to $\alpha$-helix. 
In addition, there are weakly populated minima corresponding 
to $\beta$-sheet structures at $\Delta\approx9$\,\AA\ and 
$\Delta\approx12$\,\AA. 

\begin{figure}[t]
\begin{center}
\epsfig{figure=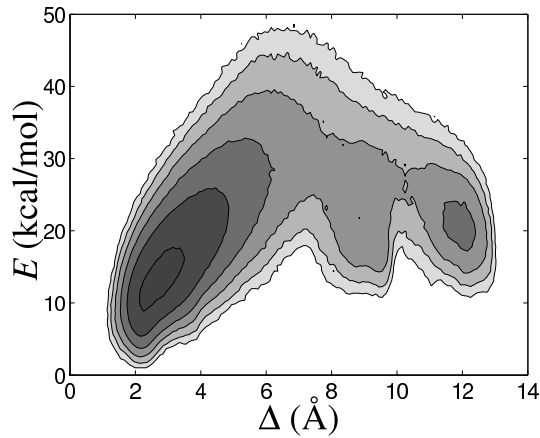,width=7cm}
\end{center}
\vspace{-3mm}
\caption{Free energy $F(\Delta,E)$ for \Fs\ at $T=284$\,K, where $\Delta$ 
denotes heavy-atom rmsd from an ideal $\alpha$-helix 
and $E$ is energy. The contours are spaced at intervals of 1\,$kT$
and dark tone corresponds to low free energy. Contours 
more than 6\,$kT$ above the minimum free energy are not shown.}
\label{fig:2}\end{figure}

\section{Results and Discussion}

\vspace{-6pt}

\subsection{$\beta$-Hairpin}

Using the model described in the previous section, we first study 
the 16-amino acid C-terminal $\beta$-hairpin from the protein G B1
domain. An important quantity monitored in our earlier study of
this peptide\cite{Irback:03} was the hydrophobicity energy $\Ehp$. 
This variable should be strongly correlated 
with Trp fluorescence, which Mu\~noz~\etal~\cite{Munoz:97} used 
to characterize the melting behavior of this peptide.  
The temperature dependence of $\Ehp$ was found to be in reasonable 
agreement with the data of Mu\~noz~\etal\ 
Several other groups have performed atomic   
simulations of the same $\beta$-hairpin, 
with\cite{Roccatano:99,Pande:99,Garcia:01,Zhou:01,Zhou:03} or
without\cite{Kussell:02,Zhou:03,Dinner:99,Zagrovic:01} explicit water. 
In contrast to ours, most models seem to require further calibration 
in order not to show a temperature dependence much weaker  
than that of experimental data.  

Figure~3a shows the temperature dependence of $\Ehp$ in 
the revised model.
\begin{figure}[t]
\begin{center}
\epsfig{figure=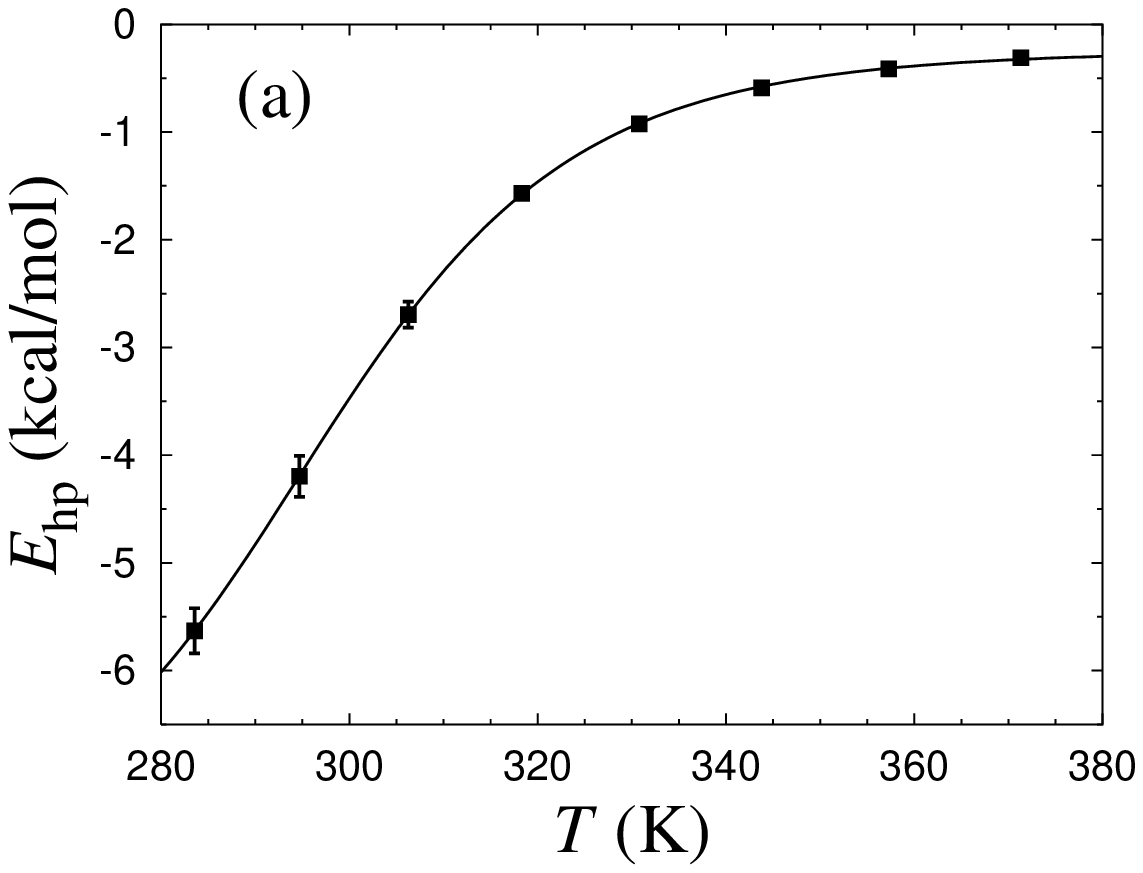,width=7cm}
\hspace{5mm}
\epsfig{figure=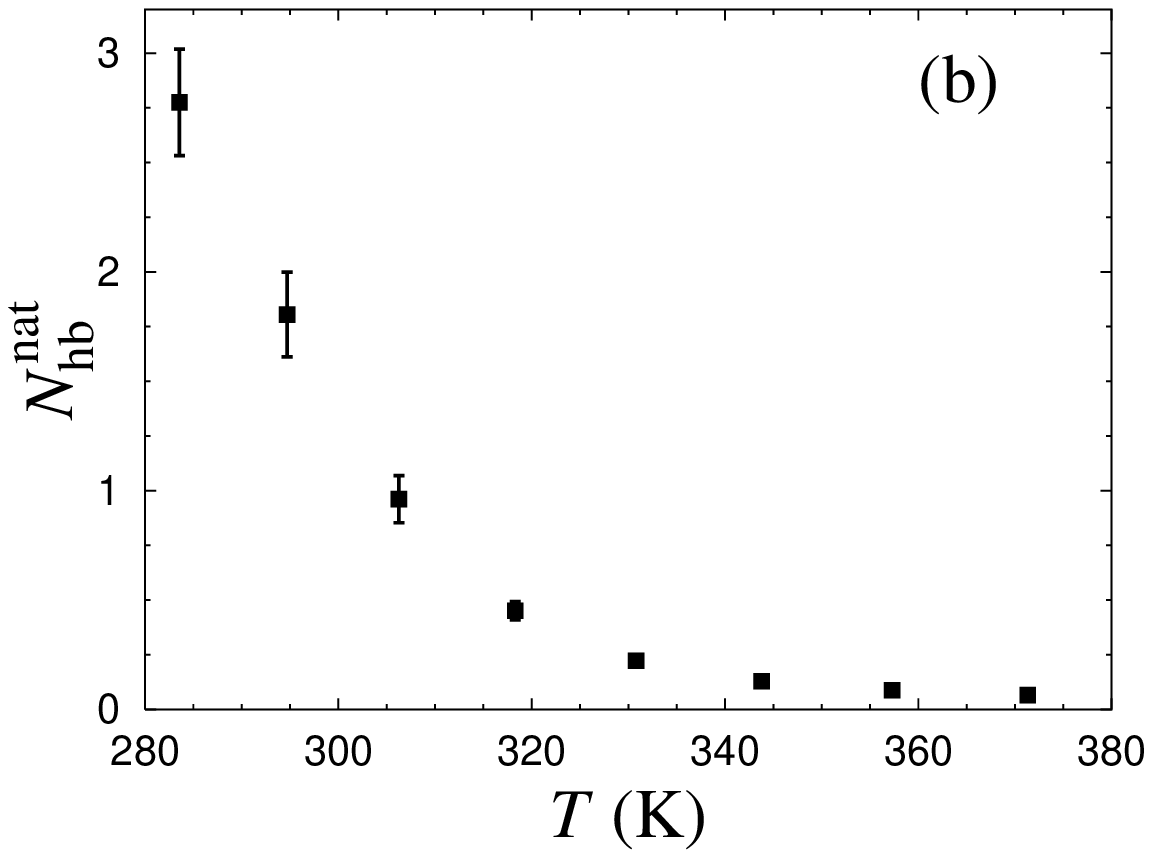,width=7cm}
\end{center}
\vspace{-3mm}
\caption{
The temperature dependence of (a) the hydrophobicity energy $\Ehp$ and 
(b) the number of native hydrogen bonds, $\Nhb$, for the $\beta$-hairpin.
The line in (a) is a fit to the two-state expression 
$\Ehp(T)=(\Ehpu+\Ehpn K(T))/(1+K(T))$, where $K(T)=\Pn(T)/\Pu(T)$, 
$\Pn(T)$ and $\Pu(T)$ being the populations of the native and unfolded
states, respectively. The effective equilibrium constant $K(T)$ is assumed
to have the first-order form $K(T)=\exp[(1/kT-1/k\Tm)\dE]$,
where $\Tm$ is the midpoint temperature and $\dE$ is the energy difference
between the two states. The baselines $\Ehpn$ and $\Ehpu$ are taken 
as constants.}
\label{fig:3}\end{figure}
The line is a fit of the data to a simple (first-order) two-state 
expression. The parameters of the fit are the midpoint temperature 
$\Tm$, the energy difference $\dE$, and two baselines. We use the 
parameter $\Tm$ to set the energy scale of the model; this parameter
is taken as $\Tm=297$\,K as determined by Mu\~noz~\etal\cite{Munoz:97} 
For the energy difference, we then obtain $\dE=13.1$\,kcal/mol. 
These values of the two-state parameters 
$\Tm$ and $\dE$ correspond to a native population 
of 74\% at $T=284$\,K, which agrees well with the result 
of Mu\~noz~\etal, 72\% at $T=284$\,K.\cite{Munoz:97} The NMR 
analysis of Blanco~\etal\cite{Blanco:94} gave, by contrast, 
a lower native population, 42\% at $T=278$\,K. A possible 
explanation of this discrepancy would be that this peptide
does not show a clear two-state behavior; the apparent native
population may then very well depend on which quantity is
studied. At first glance, this explanation may seem unlikely,   
given that the temperature dependence of the Trp fluorescence data 
to a good approximation showed two-state character.\cite{Munoz:97} 
Let us therefore stress that, despite that the two-state fit 
in Fig.~3a looks quite good, this sequence does not show ideal 
two-state behavior in our model. 
This can be seen, for example, from the energy distribution, 
which lacks a clear bimodal shape. 
This was shown in our earlier study,\cite{Irback:03} and holds true in the 
revised model as well. Similar results have also been obtained in 
simulations of a designed, fast-folding three-helix-bundle 
protein.\cite{Favrin:03}  
    
In Fig.~3b we show the temperature dependence of
the number of native hydrogen bonds, $\Nhb$, which we expect to be  
more strongly correlated than $\Ehp$ with the NMR measurements
of Blanco~\etal\ 
For $\Nhb$, a two-state fit is not meaningful; for that, 
further data at lower temperatures would be needed. 
On the other hand, the quantity $\Nhb$ can be used as a 
direct measure of nativeness. Based on inspection of many examples,
we use as a criterion for nativeness that at most two of the
native hydrogen bonds should be missing, which can be used for
the two other peptides too (see below). For the $\beta$-hairpin
with seven native hydrogen bonds (see Fig.~1a), this criterion
($\Nhb\ge5$) gives a native population of 39\% at $T=284$\,K. 
This value is close to the estimate of Blanco~\etal,\cite{Blanco:94}
42\% at $T=278$\,K. Due to uncertainties about the precise definitions
of nativeness and of when a hydrogen bond is formed, this agreement
could be somewhat accidental. There is no doubt, however, that      
the native population obtained using $\Nhb$ is significantly lower 
than that obtained above using $\Ehp$. Figure~4 shows the probability
distributions of $\Nhb$ at $T=284$\,K and $T=306$\,K. The number 
of native hydrogen bonds is seen to rapidly decrease with increasing $T$, 
as it should.

\begin{figure}[t]
\begin{center}
\epsfig{figure=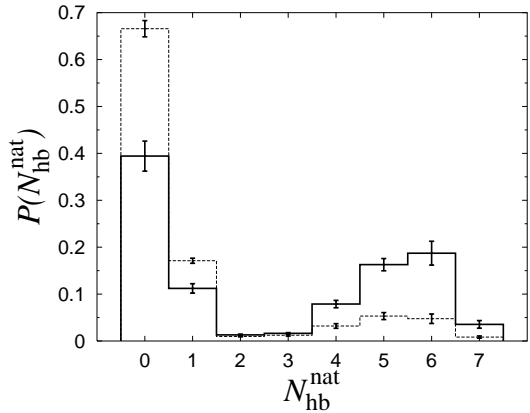,width=7cm}
\end{center}
\vspace{-3mm}
\caption{Histogram of the number of native hydrogen bonds, $\Nhb$, at 
$T=284$\,K (full line) and $T=306$\,K (dashed line) for the $\beta$-hairpin.}  
\label{fig:4}\end{figure}

The two-state parameter $\dE$ extracted from the $\Ehp$ data  
is somewhat smaller here, $\dE=13.1$\,kcal/mol, than it was in our 
earlier study, $\dE=16.1$\,kcal/mol.\cite{Irback:03} 
The reason for this is not so much that the model has changed, 
but rather that the fits were done in different ways. 
In our previous study, $\Tm$ was held fixed 
at the specific heat maximum. Here, following the analysis of
Mu\~noz~\etal\ more closely, we take $\Tm$ to be a parameter 
of the fit. The fitted value of $\Tm$ turns out to lie slightly 
below (1--2\%) the specific heat maximum. Our new analysis improves
the agreement with the result of Mu\~noz~\etal, which was
$\dE=11.6$\,kcal/mol.\cite{Munoz:97}    

Although the precise shape of the structures with lowest energy is 
sensitive to the details of the model, it is also interesting
to make an rmsd-based comparison with experimental data. 
For this purpose, we use the NMR structure for the full 
protein G B1 domain
(PDB code 1GB1, first model),\cite{Gronenborn:91}  
as the NMR restraints 
for the isolated $\beta$-hairpin were insufficient to determine a unique 
structure. Figure~5a shows the free energy $F(\Delta,E)$  
calculated as a function of rmsd, $\Delta$, and energy, $E$, at $T=284$\,K.  
\begin{figure}[t]
\begin{center}
\epsfig{figure=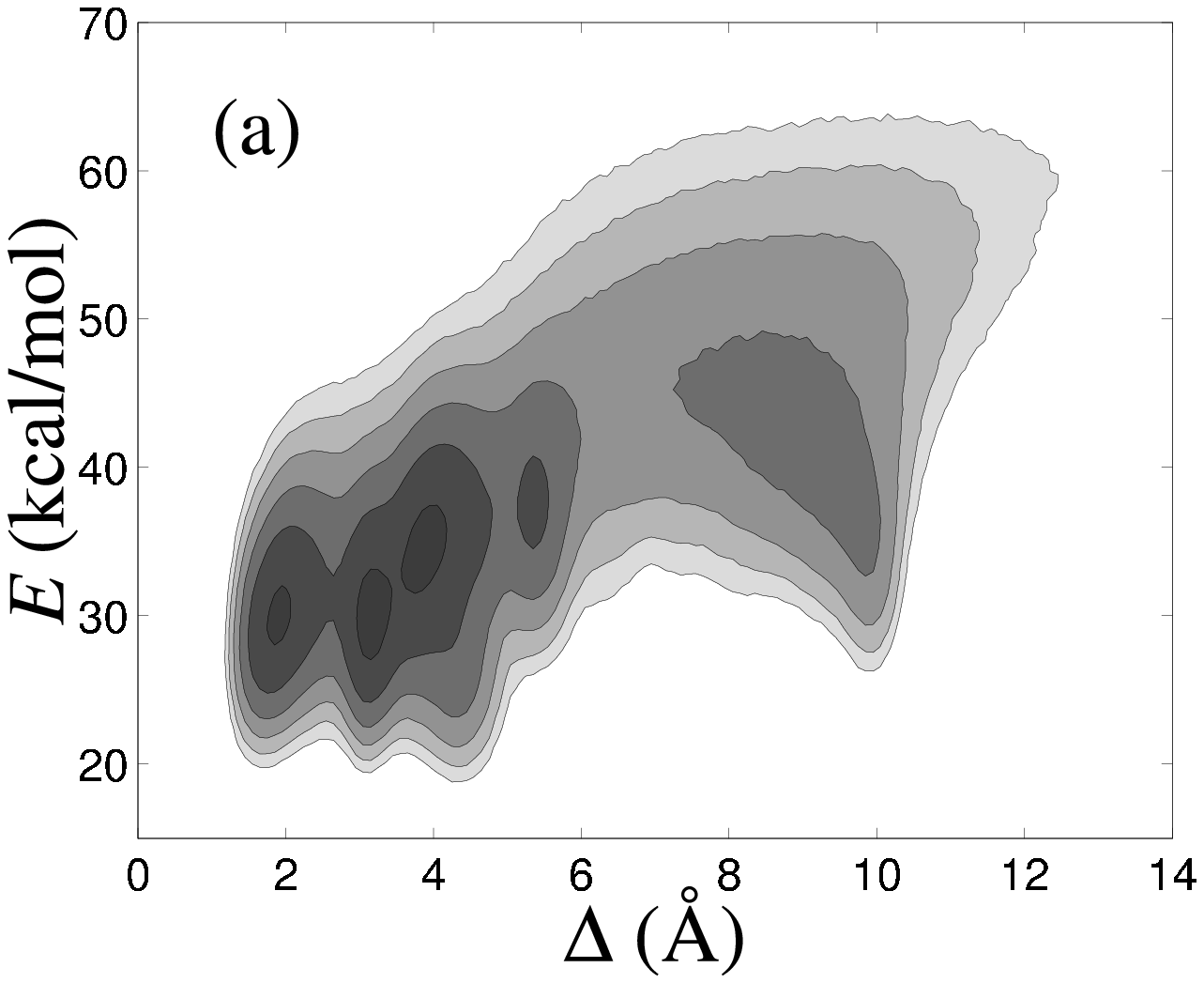,width=7cm}
\hspace{5mm}
\epsfig{figure=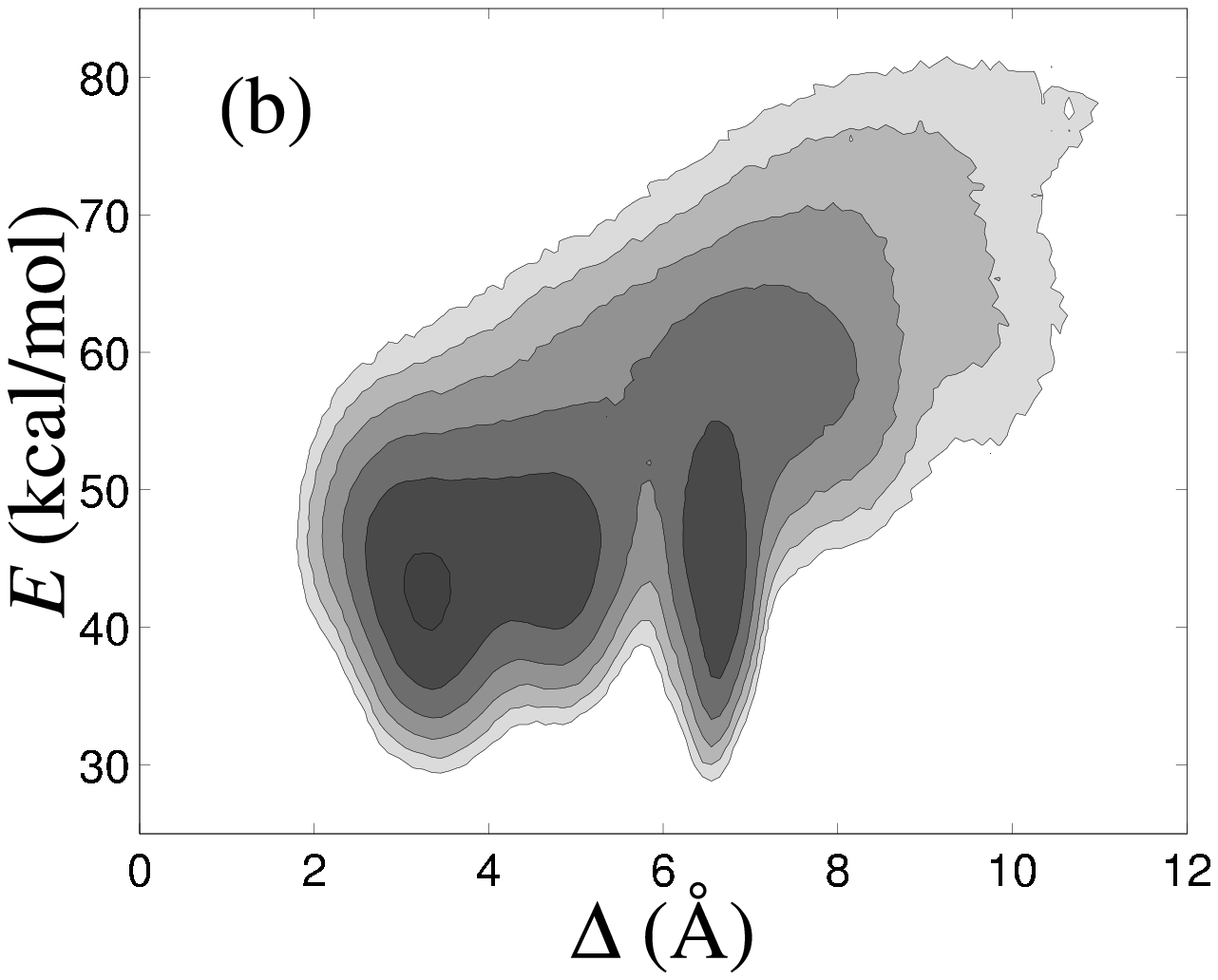,width=7cm}
\end{center}
\vspace{-3mm}
\caption{
Free energy $F(\Delta,E)$ at $T=284$\,K for (a) 
the $\beta$-hairpin and (b) the peptide LLM. $E$ is energy and
$\Delta$ is a heavy-atom rmsd, calculated using all the 
16 amino acids for the $\beta$-hairpin and amino acids 3--18 for 
LLM. The first two and last two amino acids of LLM 
do not take part in the $\beta$-sheet structure.
The contour levels are as in Fig.~2.}
\label{fig:5}\end{figure}
Three distinct, highly populated minima can be seen. 
The two minima with lowest $E$ are found at $\Delta\approx2.0$\AA\ and 
$\Delta\approx3.1$\AA, respectively. Both these correspond to 
$\beta$-hairpin structures with a high $\Nhb$. That $\Nhb$ is high 
implies, in particular, that the topology of the $\beta$-hairpin is 
the native one. The main difference between these two minima lies in the shape 
of the turn. The third minimum, at $\Delta\approx4.0$\AA, is 
somewhat higher in $E$ than the first two. This minimum is
also dominated by $\beta$-hairpin structures with the native 
topology and many hydrogen bonds, but the two strands tend to be out of
register with each other, so $\Nhb$ is low. 
Largely, it is the existence of this third minimum 
that makes the apparent native population depend on which of the 
observables $\Ehp$ and $\Nhb$ we use. Finally, there are also
two weakly populated free-energy minima corresponding to $\beta$-sheet
structures with the non-native topology ($\Delta\approx5.3$\,\AA) and
$\alpha$-helix ($\Delta\approx8$--10\,\AA), respectively. 

\subsection{Three-Stranded $\beta$-Sheets}

The {\it de novo} design of the 20-amino acid three-stranded antiparallel 
$\beta$-sheet peptide Betanova was reported in 1998.\cite{Kortemme:98}  
Recently, mutants of this peptide with higher stability were 
created by \lop~\etal\cite{Lopez:01} Among the most stable mutants found 
was the triple mutant LLM (Val5Leu, Asn12Leu, Thr17Met). 
The peptide LLM and the original Betanova were estimated\cite{Lopez:01}  
to have native populations of 36\% and 9\%, respectively, at
$T=283$\,K, based on NMR data. Melting curves have, as far as we know, 
not been reported for these peptides.  
 
Our simulations of LLM show first of all that this sequence does make
a three-stranded antiparallel $\beta$-sheet in this model. This can be seen 
from Fig.~5b, which shows the free energy $F(\Delta,E)$ at 
$T=284$\,K. The free energy has a broad minimum at
$\Delta\approx3$--5\,\AA, corresponding to $\beta$-sheet structures with 
the native topology and a high $\Nhb$. The shape of the
$\beta$-sheet varies within the minimum. At $\Delta\approx3.4$\,\AA, 
where the free energy is lowest, the $\beta$-sheet has a bent 
shape, which enables the chain to make strong hydrophobic contacts. 
At $\Delta\approx4.5$\,\AA, the $\beta$-sheet tends to be much flatter, 
which is hydrophobically disfavored but makes it possible for the 
chain to form more perfect hydrogen bonds. There is also a 
free-energy minimum at $\Delta\approx6.5$\,\AA, which  
corresponds to three-stranded antiparallel $\beta$-sheet structures 
with the non-native topology. However, the native topology is the
thermodynamically favored one. Note that the native and 
non-native topologies exhibit non-overlapping sets of backbone-backbone 
hydrogen bonds, so $\Nhb$ is low at the $\Delta\approx6.5$\,\AA\ minimum.  

The main reason why the model favors the native topology over the non-native 
one lies in the side-chain orientations    
for the hydrophobic pairs Trp3-Leu12 and Leu5-Tyr10. The \Ca-\Cb\
vectors of these pairs point inwards in the non-native topology, which makes 
it difficult to achieve proper contacts between the side chains. 
This is much easier to accomplish in the native topology, 
where the \Ca-\Cb\ vectors point outwards. 
Interestingly, the situation is similar for the $\beta$-hairpin 
above.\cite{Irback:03} The $\beta$-hairpin also has two 
pairs of hydrophobic side chains that are `bow-legged' 
in the native topology and `knock-kneed' in the non-native one.

Next we estimate the native population for LLM. As
we want to compare with the NMR-based results of \lop~\etal,
we consider $\Nhb$ rather than $\Ehp$. Figure~6a shows 
the $\Nhb$ distribution at $T=284$\,K. In addition to the native and
non-native peaks at high and low $\Nhb$, respectively, this 
distribution exhibits a third peak at $\Nhb=4$. The typical
conformation at this peak contains only the first of the two native 
$\beta$-turns (see Fig.~1b). The second $\beta$-turn is 
less stable, as will be discussed below. Using the criterion that at most 
two native hydrogen bonds should be missing ($\Nhb\geq6$), we obtain
a native population of 38\% at $T=284$\,K for LLM,
which agrees well with the result of \lop~\etal,\cite{Lopez:01}
36\% at $T=283$\,K. We also performed simulations of the original Betanova, 
and Fig.~6a shows the result for this sequence too. 
From this figure it is evident that Betanova is less stable 
than LLM. The probability that $\Nhb\geq6$ is 14\% for 
Betanova at $T=284$\,K, which means that this criterion gives
a native population close to the NMR-based result of 
\lop~\etal~\cite{Lopez:01} not only for LLM but also for Betanova.

\begin{figure}[t]
\begin{center}
\epsfig{figure=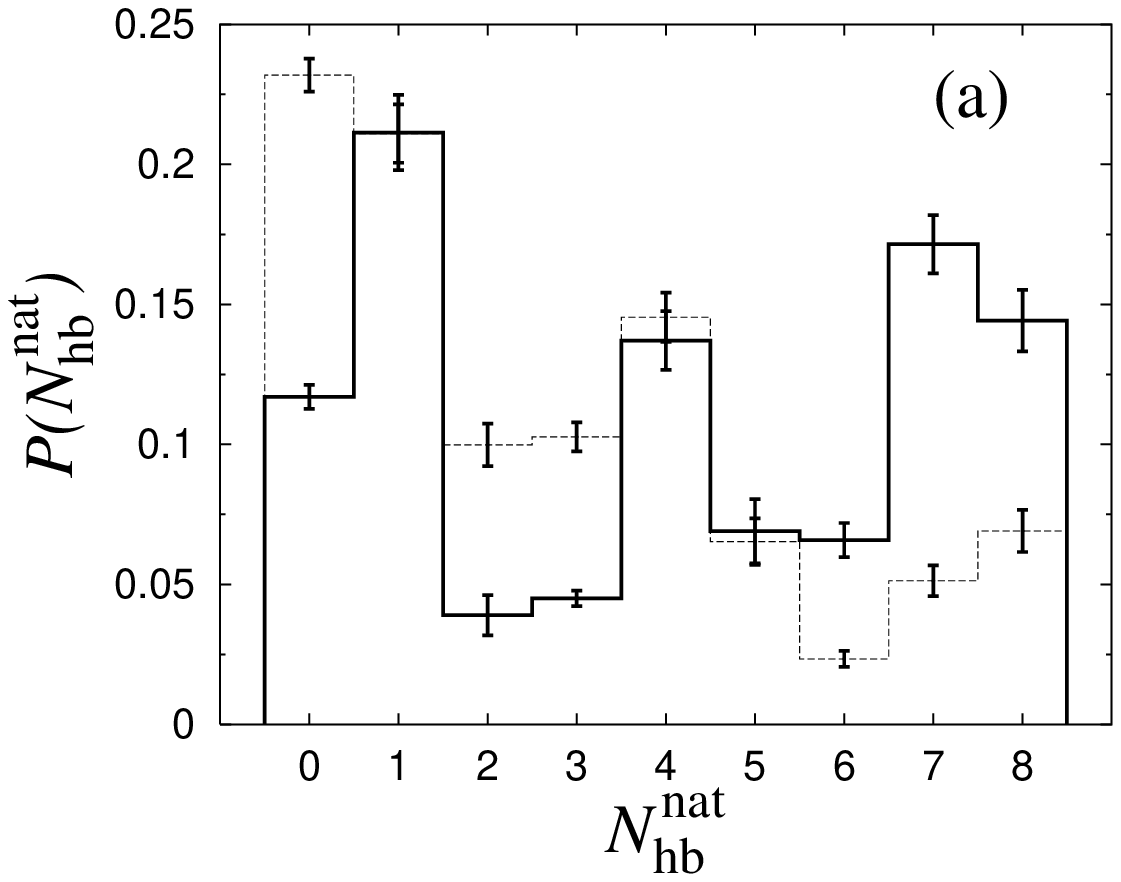,width=7cm}
\hspace{5mm}
\epsfig{figure=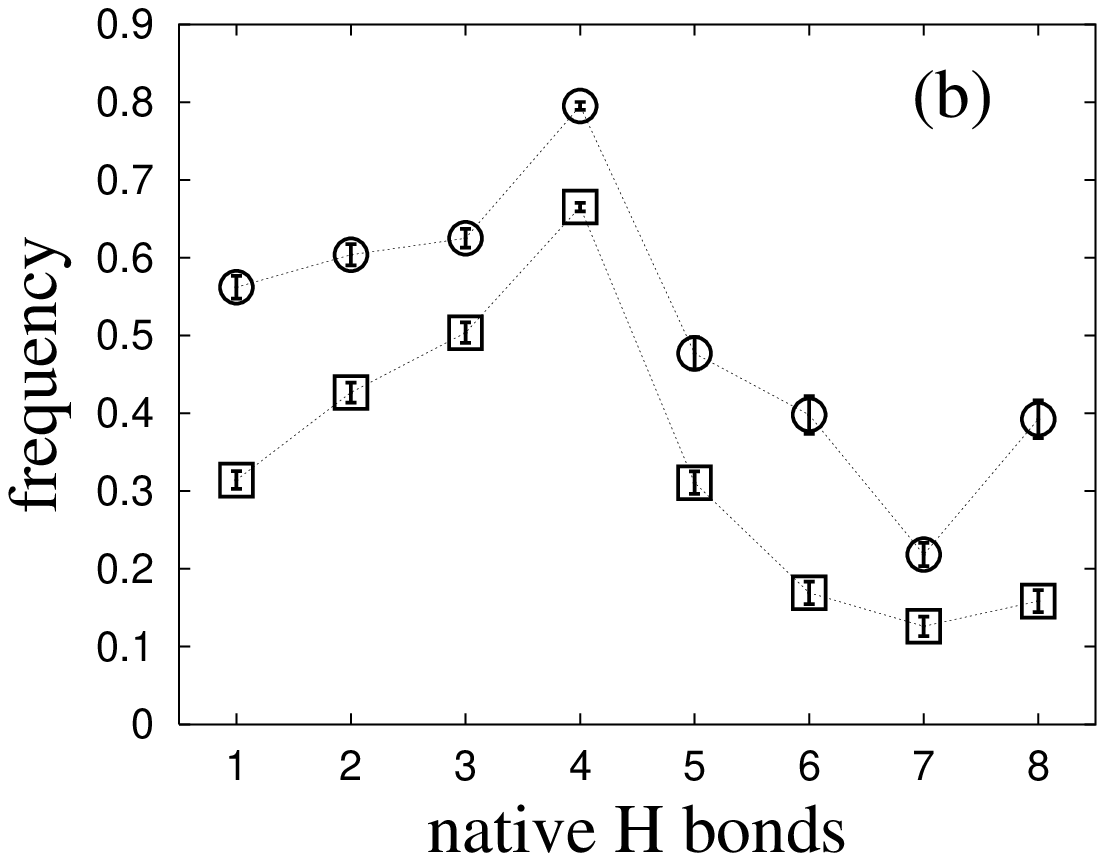,width=7cm}
\end{center}
\vspace{-3mm}
\caption{
(a) Histogram of the number of native hydrogen bonds, $\Nhb$, 
at $T=284$\,K for LLM (full line) and Betanova (dashed line).
(b) The frequency of occurrence  
for the eight native hydrogen bonds (labeled according to  
Fig.~1b) for LLM ($\circ$) 
and Betanova ($\Box$) at $T=284$\,K.}
\label{fig:6}\end{figure}

That the model predicts LLM to be more stable than
Betanova is not surprising because LLM has a more pronounced 
hydrophobic core. The agreement with experimental data is, nevertheless,  
remarkably good, especially since these calculations do not involve
any adjustable parameter; the energy scale of the model is 
fixed using melting data for the $\beta$-hairpin and is then 
left unchanged. 

Figure~6b shows the frequencies of occurrence for the 
different native hydrogen bonds (see Fig.~1b)
for LLM and Betanova. For Betanova, there is a clear difference
between the hydrogen bonds involved in the first $\beta$-turn
(1--4) and those involved in the second $\beta$-turn (5--8).
The latter four occur infrequently, showing that the 
second $\beta$-turn is quite unstable, which is in line with 
the conclusions of \lop~\etal\cite{Lopez:01}     
For LLM, the difference in stability between the two $\beta$-turns
is less pronounced. However, hydrogen bond 7, which connects Met17 to   
Tyr10 (see Fig.~1b), is quite unstable.
The reason for this is that the side chain of Met17 can make 
better contacts with other hydrophobic side chains if the 
strand is slightly bent. This bend makes it difficult 
for hydrogen bond 7 to form. 

Finally, in Fig.~7 we show the temperature dependence of $\Ehp$ 
and $\Nhb$ for LLM. As in the $\beta$-hairpin case, we find that 
a simple two-state fit provides a good description of the data for 
$\Ehp$. The fitted values of the parameters $\Tm$ and $\dE$  
are $\Tm=303$\,K and $\dE=13.0$\,kcal/mol, which means that the native 
population obtained from this fit is significantly higher than that 
obtained from the $\Nhb$ distribution (see Fig.~6a). 
So, the model predicts that the apparent native population depends on 
which observable is used for this sequence, too.  We are not aware of 
any existing experimental data that support, or refute, this conclusion
for LLM.   

\begin{figure}[t]
\begin{center}
\epsfig{figure=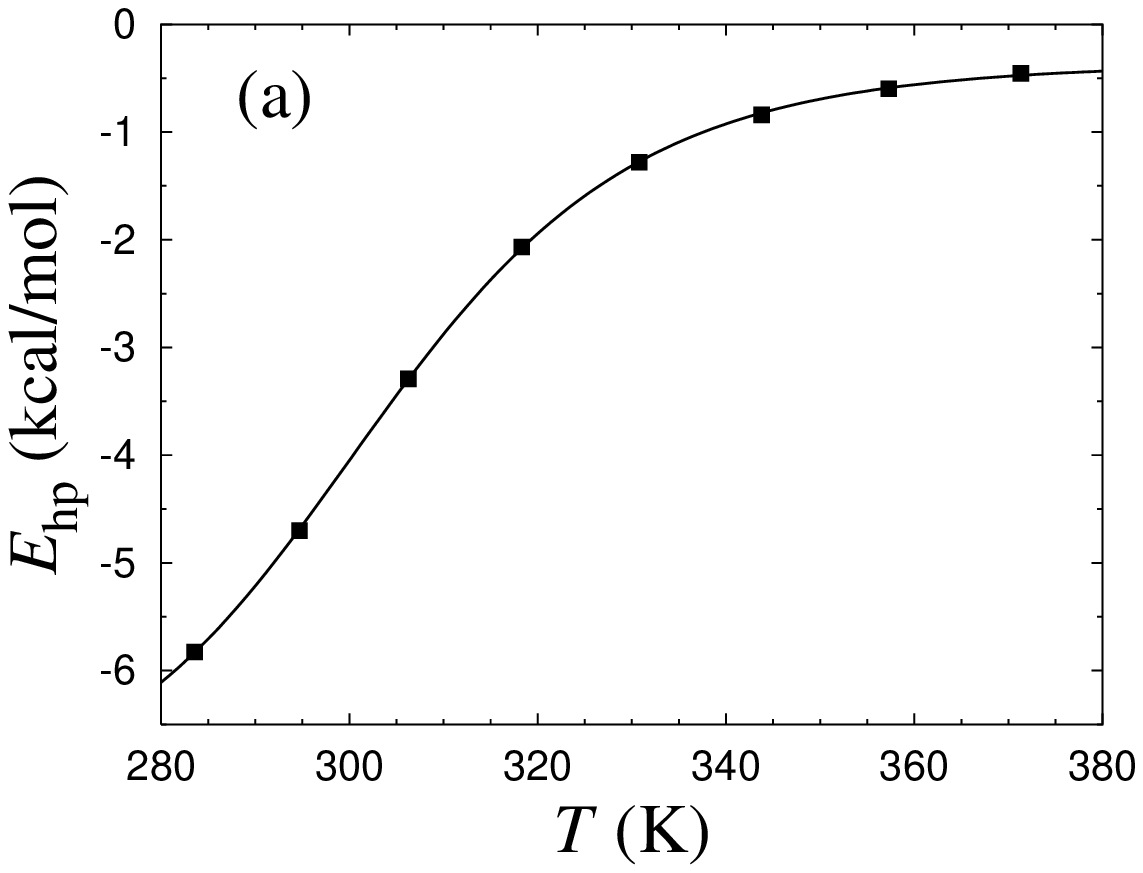,width=7cm}
\hspace{5mm}
\epsfig{figure=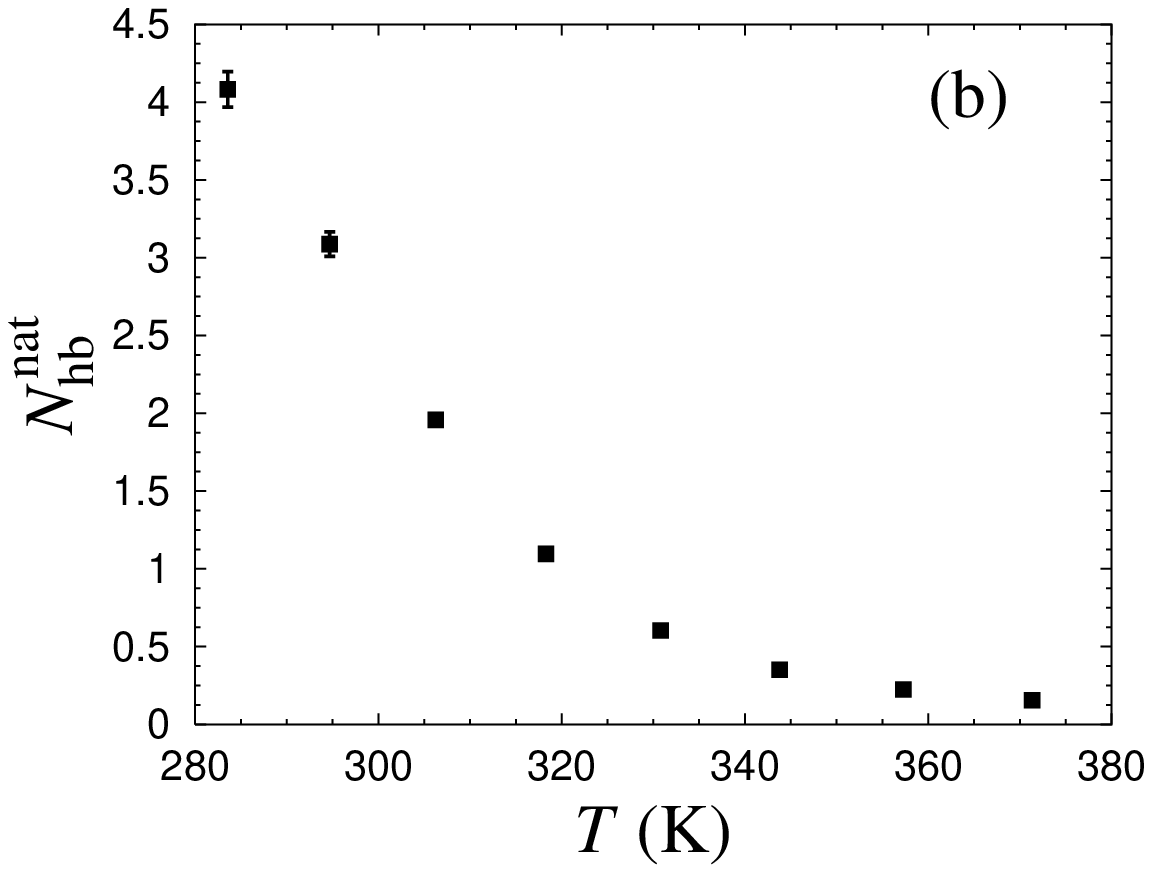,width=7cm}
\end{center}
\vspace{-3mm}
\caption{
The temperature dependence of (a) the hydrophobicity energy $\Ehp$ and 
(b) the number of native hydrogen bonds, $\Nhb$, for the LLM peptide.
The line in (a) is a first-order two-state fit  
(as in Fig.~3).}
\label{fig:7}\end{figure}

\newpage
    
\section{Conclusion}

Using a novel all-atom model with a simplified sequence-based potential,
we have investigated the equilibrium behaviors of three $\beta$-sheet 
peptides. We determined native populations for these peptides 
in two ways, from the distribution of the number of native hydrogen bonds 
($\Nhb$) and from the temperature dependence of the hydrophobicity energy 
($\Ehp$). These two estimates were compared with experimental results 
based on NMR and Trp fluorescence, respectively. This comparison is 
summarized in Table~II. The agreement with experimental data is good,
which in particular means that the model to a good approximation is
able to reproduce the relative stabilities of these three peptides, as 
obtained from the NMR measurements. In line with the experimental
results on the $\beta$-hairpin, we find 
that the apparent native population depends on whether we use 
$\Nhb$ or $\Ehp$.  This reflects the fact that the melting 
transition is not a clear two-state transition in our model (for any of
these three sequences). It is also worth noting that, despite that the 
two-state picture is an oversimplification, the temperature dependence 
of $\Ehp$ is quite well described by a simple two-state expression 
(see Figs. 3a and 7a). Computational 
studies of the $\beta$-hairpin have also been performed by many 
other groups, but the temperature dependence obtained was typically 
too weak, as has been pointed out by Zhou~\etal~\cite{Zhou:01} Our
model shows a temperature dependence which is in good agreement with 
experimental data. 

\begin{table}[t]
\begin{center}
\begin{tabular}{llllll}
&\multicolumn{2}{c}{Model, 284\,K}&& 
\multicolumn{2}{c}{Experiment} \\
\cline{2-3}\cline{5-6}
& $\Nhb$ & $\Ehp$ && NMR & Trp fluorescence\\
\hline
$\beta$-hairpin & 39\% & 74\% &\ \ & 
42\%, 278\,K~\protect\cite{Blanco:94} & 72\%, 284\,K~\protect\cite{Munoz:97}\\
LLM             & 38\% && & 36\%, 283\,K~\protect\cite{Lopez:01} &\\
Betanova        & 14\% && & 9\%, 283\,K~\protect\cite{Lopez:01} &
\end{tabular}
\caption{
Summary of apparent native populations obtained from 
simulations and experimental data, respectively (see the text). 
The model results have statistical errors of 1--4\%. 
}
\label{tab:2}
\end{center}
\end{table}

Our study of these three different peptides was carried out using one
and the same set of parameters. In addition, we showed that the \Fs\ peptide 
makes an $\alpha$-helix for this choice of parameters. While these results
are very encouraging, it is important to stress 
that we do not expect the model to be directly applicable 
to other sequences. However, by confronting the model with new sequences,
we hope it will be possible to refine the potential, and thereby 
further extend its applicability. The present study was a first step
in this direction, in which the model was improved by studying
LLM and Betanova. To make the model able to fold these sequences, 
many changes were made, several of which were minor. The two perhaps 
most important changes were the replacement of the old hydrophobicity 
matrix ($\Mij$), and the introduction of a simple form of context 
dependence for the hydrogen bonds. Whether it will be possible to carry 
on this process to a point where the model correctly reproduces the
thermodynamics of small proteins remains to be seen. One thing that 
probably will be necessary in order to achieve this goal is to include
multibody effects in the hydrophobicity potential; the present pairwise
additive potential is likely to become insufficient as the chains get
larger. Computationally, there is room for extending the calculations
to larger chains; the calculations presented here required about     
two weeks on a standard desktop computer for each peptide.  

\vspace{12pt}

\noindent
{\bf Acknowledgments:}
We thank Luis Serrano and Manuela L\'opez de la Paz for providing
NMR data for LLM and Betanova. 
This work was in part supported by the Swedish Foundation for Strategic 
Research and the Swedish Research Council.


\newpage

\begin{thebibliography}{}

\bibitem{Kortemme:98}
Kortemme T, Ram\'\i rez-Alvarado M, Serrano L.
Design of a 20-amino acid, three-stranded $\beta$-sheet protein. 
\Sci 1998;\,281:\,253--256.

\bibitem{Lopez:01}
L\'opez de la Paz M, Lacroix E, Ram\'\i rez-Alvarado M, Serrano L.
Computer-aided design of $\beta$-sheet peptides.
\JMB 2001;\,312:\,229--246.

\bibitem{deAlba:99}
de Alba E, Santaro J, Rico M, Jim\'enez MA.
De novo design of a monomeric three-stranded antiparallel $\beta$-sheet.
\ProSci 1999;\,8:\,854--865.

\bibitem{Bursulaya:99}
Bursulaya BD, Brooks CL III.
Folding free energy surface of a three-stranded $\beta$-sheet protein.
\JACS 1999;\,121:\,9947--9951.

\bibitem{Colombo:02}
Colombo C, Roccatano D, Mark AE.
Folding and stability of the three-stranded $\beta$-sheet peptide betanova:
Insights from molecular dynamics simulations. 
\Pro 2002;\,46:\,380--392.

\bibitem{Cavalli:03}
Cavalli A, Haberth\"ur U, Paci E, Caflisch A.
Fast protein folding on downhill energy landscape.
\ProSci 2003;\,12:\,1801--1803.

\bibitem{Karanicolas:03}
Karanicolas J, Brooks CL III.
The structural basis for biphasic kinetics in the folding of the WW
domain from a formin-binding protein: Lessons for protein design?
\PNAS 2003;\,100:\,3954--3959.

\bibitem{Granakaran:03}
Granakaran S, Nymeyer H, Portman J, Sanbonmatsu KY, Garc\'\i a AE.
Peptide folding simulations.
\COSB 2003;\,13:\,168--174.

\bibitem{Irback:03}
Irb\"ack A, Samuelsson B, Sjunnesson F, Wallin S.
Thermodynamics of $\alpha$- and $\beta$-structure formation in proteins.
\BJ 2003;\,85:\,1466--1473.

\bibitem{Lockhart:92}
Lockhart DJ, Kim PS.
Internal Stark effect measurement of the electric field
at the amino terminus of an $\alpha$ helix.
\Sci 1992;\,257:\,947--951.
 
\bibitem{Lockhart:93}
Lockhart DJ, Kim PS.
Electrostatic screening of charge and dipole interactions
with the helix backbone.
\Sci 1993;\,260:\,198--202.

\bibitem{Kussell:02}
Kussell E, Shimada J, Shakhnovich EI. 
A structure-based method for derivation of all-atom potentials for
protein folding.
\PNAS 2002;\,99\,:5343--5348.

\bibitem{Munoz:97}
Mu\~noz V, Thompson PA, Hofrichter J, Eaton WA.
Folding dynamics and mechanism of $\beta$-hairpin formation.
\Nat 1997;\,390:\,196--199.

\bibitem{Branden:91}
Branden C, Tooze J. {\it Introduction to Protein Structure}.
New York: Garland Publishing; 1991.

\bibitem{Miyazawa:96}
Miyazawa S, Jernigan RL.
Residue-residue potentials with a favorable contact pair term
and an unfavorable high packing density, for simulation and
threading.
\JMB 1996;\,256:\,623--644.

\bibitem{Lyubartsev:92}
Lyubartsev AP, Martsinovski AA, Shevkunov SV, 
Vorontsov-Velyaminov PN.
New approach to Monte Carlo calculation of the free energy:
Method of expanded ensembles.
\JCP 1992;\,96:\,1776--1783.

\bibitem{Marinari:92}
Marinari E, Parisi G.
Simulated tempering: A new Monte Carlo scheme.
\EL 1992;\,19:\,451--458.

\bibitem{Irback:95}
Irb\"ack A, Potthast F.
Studies of an off-lattice model for protein folding: Sequence
dependence and improved sampling at finite temperature.
\JCP 1995;\,103:\,10298--10305.

\bibitem{Hansmann:99}
Hansmann UHE, Okamoto Y.
New Monte Carlo algorithms for protein folding.
\COSB 1999;\,9:\,177--183.

\bibitem{Lal:69}
Lal M.
Monte Carlo computer simulation of chain molecules. I.
\MP 1969;\,17:\,57--64. 

\bibitem{Favrin:01}
Favrin G, Irb\"ack A, Sjunnesson F.
Monte Carlo update for chain molecules: Biased Gaussian steps
in torsional space.
\JCP 2001;\,114:\,8154--8158.

\bibitem{NR}
Press WH, Flannery BP, Teukolsky SA, Vetterling WT.
{\it Numerical Recipes in C: The Art of Scientific Computing}.
Cambridge: Cambridge University Press; 1992. 

\bibitem{Blanco:94}
Blanco FJ, Rivas G, Serrano L.
A short linear peptide that folds into a native stable
$\beta$-hairpin in aqueous solution.
\NSB 1994;\,1:\,584--590.

\bibitem{Roccatano:99}
Roccatano D, Amadei A, Di Nola A, Berendsen HJC.
A molecular dynamics study of the 41--56 $\beta$-hairpin
from B1 domain of protein G.
\ProSci 1999;\,8:\,2130--2143.

\bibitem{Pande:99}
Pande VS, Rokhsar DS.
Molecular dynamics simulations of unfolding and refolding
of a $\beta$-hairpin fragment from protein G.
\PNAS 1999;\,96:\,9062--9067.

\bibitem{Garcia:01}
Garc\'\i a AE, Sanbonmatsu KY.
Exploring the energy landscape of a $\beta$ hairpin in
explicit solvent.
\Pro 2001;\,42:\,345--354.

\bibitem{Zhou:01}
Zhou R, Berne BJ, Germain R.
The free energy landscape for $\beta$ hairpin folding in explicit water.
\PNAS 2001;\,98:\,14931--14936.

\bibitem{Zhou:03}
Zhou R.
Free energy landscape of protein folding in water: Explicit vs. implicit 
solvent.
\Pro 2003;\,53:\,148--161.

\bibitem{Dinner:99}
Dinner AR, Lazaridis T, Karplus M.
Understanding $\beta$-hairpin formation.
\PNAS 1999;\,96:\,9068--9073.

\bibitem{Zagrovic:01}
Zagrovic B, Sorin EJ, Pande V.
$\beta$-hairpin folding simulations in atomistic detail
using an implicit solvent model.
\JMB 2001;\,313:\,151--169.

\bibitem{Favrin:03}
Favrin G, Irb\"ack A, Samuelsson B, Wallin S.
Two-state folding over a weak free-energy barrier.
\BJ 2003;\,85:\,1457--1465.

\bibitem{Gronenborn:91}
Gronenborn, AM, Filpula DR, Essig NZ, Achari A,
Whitlow M, Wingfield PT, Clore GM.
A novel, highly stable fold of the immunoglobulin-binding
domain of streptococcal protein G.
\Sci 1991;\,253:\,657--661.

\end{thebibliography}
\end{document}